\documentclass[a4paper,11pt]{article}
\usepackage{indentfirst}
\usepackage{graphicx,epsfig,url,xcolor,hyperref}
\usepackage{graphicx}
\usepackage{dcolumn}
\usepackage{epstopdf}
\usepackage{fixmath}
\usepackage{amsmath}
\usepackage{textcomp}
\usepackage{authblk}
\usepackage[a4paper, total={6in, 8in}]{geometry}
\usepackage{cite}
\usepackage[normalem]{ulem}
\usepackage{caption}
\usepackage{subcaption}

\title{ Unimodular Theory of Gravity in Light of the Latest Cosmological Data}
\author[1]{Naveen K. Singh}
\affil[1]{Sir P.T. Sarvajanik College of Science, Surat 395001, Gujarat, India}
\author[2]{Gopal Kashyap \thanks{\href{mailto:gplkumar87@gmail.com}{gplkumar87@gmail.com}}}
\affil[2]{Department
of Physics, School of Advanced Sciences, Vellore Institute of Technology, Vellore, Tamil Nadu 632014, India}
\date{}
\begin{document}

\vspace{10mm}

\maketitle

\begin{abstract}
 The unimodular theory of gravity is an alternative perspective to traditional Einstein's general relativity and opens new possibilities for exploring its implications in cosmology. In this paper, we investigate the unimodular gravity (UG) with the latest cosmological data from the Pantheon sample of Type Ia supernovae (SN), Baryon Acoustic Oscillations (BAO), and the observational H(z) data from Differential Age method (DA). We consider a model consisting of a generalized cosmological constant with radiation and dark matter. The considered theory respects only unimodular coordinate transformations. We fit our model with low-redshift data from SN and DA and determine the value of parameter $\xi$ of the theory. We find the best-fit value of parameter $\xi =6.23 \pm 0.5$; which deviates from 6, for which the theory becomes the standard general theory of relativity. We further study the Hubble constant problem by combining the SN and DA data with BAO data. We observe deviation in the value of $H_0$ from the standard $\Lambda$CDM model.  We obtain $H_0$ as $70.7 \pm 4.1 \ \mbox{Km s}^{-1} \mbox{Mpc} ^{-1}$ and $69.24 \pm 0.90 \  \mbox{Km s}^{-1} \mbox{Mpc} ^{-1}$ from supernovae data and BAO data, respectively in unimodular gravity. Combining the BAO data with SN+DA data set, we obtain $H_0$ as $70.57 \pm 0.56 \ \mbox{Km s}^{-1} \mbox{Mpc} ^{-1}$.
\end{abstract}

\section{Introduction}
There are many competing models to fit the cosmological observations, such as the modified theory of gravity, the scalar field theory, the $\Lambda$CDM model, etc \cite{Copeland:2006wr}. In these models, the $\Lambda$CDM model is a plausible model for explaining current cosmological observations \cite{Carroll_2001,Peebles_2003}. However, the cosmological constant $\Lambda$ has its own fine tuning problem. Some of the proposals, such as supersymmetry, supergravity, anthropic considerations, adjustment mechanism, changing gravity, etc., have been discussed in Ref. \cite{Weinberg:1988cp} regarding the cosmological constant problem. In this review paper \cite{Weinberg:1988cp}, unimodular gravity (UG) is discussed as a possible generalisation of
gravity. Unimodular gravity is an interesting model that was first proposed by Anderson and Finkelstein \cite{Anderson:1971pn} following the closely related proposal given by Einstein \cite{Einstein}. It might solve the cosmological constant problem since $\sqrt{-g}$ is not a dynamical field in this theory. However, in Ref. \cite{Weinberg:1988cp}, it is explained that the cosmological constant reappears as an integration constant, and the fine tuning problem remains in the theory. This is because, within that unimodular model of gravity, the theory still maintains full general coordinate invariance (GCI). Several studies regarding the cosmological constant problem in unimodular gravity have been done in Refs. \cite{Henneaux:1989zc, Ng:1990xz, Finkelstein:2000pg, Smolin:2009ti, Jirousek:2023gzr}.  \\

In Refs. \cite{Zee,Buchmuller:1988wx,Unruh:1988in,Jain:2011jc,Jain:2012gc, Alvarez:2005iy,Alvarez:2007nn, Gao2014,Nojiri:2016ppu,Bamba:2016wjm, Rajabi2017,Rajabi2021,Costantini:2022nof,Cho:2014taa,Singh:2012sx,Odintsov:2016imq,Cedeno2021,Agrawal2023}, various other aspects of unimodular gravity have been studied. In this work, we are interested in studying its implications in cosmology. The full metric can be broken in a scalar field and a unimodular metric \cite{Zee,Buchmuller:1988wx}. In  Refs. \cite{Jain:2011jc,Jain:2012gc}, the implication of unimodular gravity has been studied thoroughly using such a decomposition. In these works and Refs. \cite{Zee,Buchmuller:1988wx}, the theory respects covariance under unimodular coordinate transformations instead of general coordinate invariance.
Broken general coordinate invariance introduces a parameter $\xi$ \cite{Jain:2011jc,Jain:2012gc} wih a value other than 6, where $\xi=6$ corresponds to full general coordinate invariance, and the outcome of the theory is the same as that of general relativity. Other values of $\xi$ correspond to unimodular gravity and provide covariance under only unimodular coordinate transformations. In Refs. \cite{Jain:2011jc,Jain:2012gc}, authors discuss the expansion of the universe considering generalized non-relativistic matter and generalized
cosmological constant separately. A model where only radiation is assumed is also discussed. The motivation for these models is to describe the current expansion of the universe by only one component, either dark matter or a cosmological constant, to solve the coincidence problem.\\

One of the challenging tasks in modern cosmology is to determine  the precise value of the Hubble constant $H_0$. Considering the standard $\Lambda$CDM model, the $H_0$ value determined by the cosmic microwave background (CMB) experiments, such as WMAP and Planck differs from the value determined by the local distance ladder approach, such as SH0ES (Supernovae and $H_0$ for the Equation of State) project. Planck 2018 results give the value, $H_0=67.4 \pm 0.5 \  \mbox{Km s} ^{-1}  \mbox{Mpc} ^{-1}$ \cite{Planck:2018vyg}. The constraint of $H_0$ in CMB measurement is model dependent. The most recent results of SH0ES program gives $H_0=73.30 \pm 1.04 \ \mbox{Km s} ^{-1} \mbox{Mpc} ^{-1}$ \cite{Riess:2021jrx}, which differ by $5 \sigma$ from the final result of Planck. Since the local measurement of $H_0$ does not rely on any cosmological assumptions, it can be considered model independent. To alleviate this tension in the measurement of $H_0$ from low and high redshift probes, primarily two methods are suggested in the literature: early universe modification and late universe modification \cite{DiValentino:2021izs, Poulin:2018cxd,Agrawal:2019lmo, Karwal:2021vpk,Cai:2021wgv}. Modified gravity might also be   a solution for the Hubble tension \cite{Renk:2017rzu,Tal:2021}.\\

In this paper, we investigate how the unimodular theory differs from the general theory of relativity and to what extent the Hubble tension problem can be addressed within this framework. Here we take into account a unimodular gravity model \cite{Jain:2012gc} with a generalized cosmological constant. We extend this model by including the dark matter and radiation energy components. We use the latest Pantheon supernovae data to estimate the parameters of the model. We further study the Hubble constant problem in the theory by including the BAO data set.

The manuscript is structured as follows. The review of the unimodular gravity model and unimodular field equations, with broken general coordinate invariance, is given in Sec. 2 and Sec. 3, respectively. In Sec. 4, we describe the unimodular gravity model having a generalized Cosmological Constant. In Sec. 5, we describe our methodology and data sets used for the analysis. In Sec. 6, we discuss our results and then we conclude in Sec. 7.
 \section{Fields Decomposition in Unimodular Gavity}
To begin, we decompose the standard metric $g_{\mu\nu}$ into a scalar field $\chi$ and unimodular metric $\bar g_{\mu\nu}$ as, $ g_{\mu\nu} = \chi^2 \bar g_{\mu \nu}$. We point out that in Cartesian coordinate system the determinant of $\bar g_{\mu\nu}$ is unity. We can generalize it to any 
 coordinate system, such that det$(\bar g_{\mu\nu}) = f(x)$ \cite{Jain:2012gc}, where $f(x)$ is some specified function of space-time coordinates while keeping det$(\bar g_{\mu\nu})$ as non dynamical. The theory is supposed to be invariant under unimodular general coordinate transformations (UGCT), such that the Jacobian of the transformation is unity, i.e.,
\begin{eqnarray}
&x^\mu \rightarrow x'^\mu \\
& \text{det} \left(\frac{\partial x'^\mu}{ \partial x^\nu} \right) =1
\end{eqnarray}
Under this transformation, the determinant of $g_{\mu\nu}$ and hence the field $\chi$ behaves as a scalar.
 Following the definition of $g_{\mu \nu}$, we have
 \begin{equation}
  g^{\mu\nu} = \frac{\bar g^{\mu\nu}}{\chi^2}, \ \ \Gamma^{\mu}_{\alpha \beta} = \bar \Gamma^{\mu}_{\alpha \beta} + \tilde{\Gamma}^{\mu}_{\alpha \beta},
 \end{equation}
where, $\Gamma^{\mu}_{\alpha \beta}$ is the full affine connection, and the connection $\bar \Gamma^{\mu}_{\alpha \beta}$ corresponds  to the unimodular metric $\bar g_{\mu\nu}$.  $\tilde{\Gamma}^{\mu}_{\alpha \beta}$ contains all the terms of the scalar field $\chi$  and is given by,
\begin{equation}
 \tilde{\Gamma}^{\mu}_{\alpha \beta} = \bar g^{\mu}_{\beta} \partial_{\alpha}  \ln \chi +  \bar g^{\mu}_{\alpha} \partial_{\beta}  \ln \chi -\bar g_{\alpha \beta} \partial^{\mu} \ln \chi .
\end{equation}
Similarly, under this definition, the Ricci curvature tensor $R_{\mu\nu}$  and Ricci scalr $R$ is decomposed. The first part of $R_{\mu \nu}$ is made of the unimodular metric and the second one contains the terms of the scalar field $\chi$ (see Ref.\cite{Jain:2012gc}). One can write the gravitational action as 
\begin{equation}
 S_{E}= \int d^4 x \frac{\sqrt{-\bar g}}{16\pi G}\Big[\chi^2 \bar R - \xi \bar g^{\mu\nu} \partial_{\mu} \chi \partial_{\nu} \chi \Big],
\end{equation}
where, if parameter $\xi \neq 6$, GCI is broken and theory respects only unimodular coordinate invariance.
\section{Field Equations in Unimodular Gravity }
We consider the action as discussed in Refs. \cite{Jain:2011jc,Jain:2012gc}. The action
follows covariance under unimodular coordinate transformations. Considering both matter and cosmological constant, the action is given as follows:
\begin{equation}
 S= \int d^4 x \sqrt{-\bar g} \Big[\frac{\chi^2}{\kappa} \bar R  - \frac{\xi}{\kappa} \bar g^{\mu\nu} \partial_{\mu} \chi \partial_{\nu} \chi \Big]  + S_{M} + S_{\Lambda}.
\end{equation}
Here, $\kappa= 16 \pi G$ and $\xi$ is the parameter of the theory, $S_{M}$ and $S_{\Lambda}$ are the actions corresponding to matter and cosmological constant. For the general theory of relativity, the parameter $\xi =6$. Under this theory, the Einstein field equation and equation of motion for field $\chi$ are given by
\begin{eqnarray}
 -\chi^2 \Big [ \bar R_{\mu\nu} - \frac{1}{4}
 \bar g_{\mu\nu} \bar R \Big] - \Big[
 (\chi^2)_{;\mu;\nu} - \frac{1}{4}
 \bar g_{\mu\nu} (\chi^2)_{;\lambda}^{;\lambda}\Big] + \xi \Big[ \partial_{\mu}\chi\partial_{\nu} \chi - \frac{1}{4}
 \bar g_{\mu\nu} \partial^{\lambda} \chi\partial_{\lambda} \chi \Big] \nonumber \\  = \frac{\kappa}{2} \Big[
 T_{\mu\nu} - \frac{1}{4}
 \bar g_{\mu\nu} T^{\lambda}_{\lambda}\Big] \label{FEQ1}
\end{eqnarray}
and,
\begin{equation}
 2 \chi \bar R +  2\xi \bar g^{\mu\nu} (\chi)_{; \mu ; \nu} = \kappa T_{\chi},  \label{FEQ2}
\end{equation}
respectively. 

In this theory the cosmological constant or vacuum energy term does not contribute to Eq. \ref{FEQ1}, however the term $T_{\chi}$ includes all the contribution due to the coupling of $\chi$ with matter fields and the cosmological constant. For the spatially flat Friedmann-Robertson-Walk (FLRW) metric we can write $\bar g_{\mu \nu} = \mbox{diagonal}[1,-1,-1,-1]$, and identify the scale factor $a(\eta)$ with the scalar field $\chi$.

\section{Generalized Cosmological Constant with Radiation}
In  Ref\cite{Jain:2012gc}, the authors investigate a model for the coincidence problem based on a generalized cosmological constant in unimodular gravity, where only ordinary matter and the cosmological constant play a role. The action for the generalized cosmological constant term is defined as,
\begin{equation}
 S_{\Lambda} = - \int d^4 x \sqrt{-\bar g} \Big[\Lambda\chi^{\xi-2} + \Lambda_1 \chi\Big] ,
\end{equation}

In this model, the generalized cosmological constant term $\Lambda \chi^\delta$ with a parameter $\delta$ appears in the action instead of the standard general relativity term $\Lambda \chi^4$. For a consistent solution, it is found that  a term $\Lambda_1 \chi$ is required in the action and $\delta=\xi-2$. We follow this model. We are interested in considering both the dark matter (CDM) and cosmological constant term, along with radiation. Thus, in place of ordinary matter, we take into account both cold dark matter and ordinary matter, and we also include radiation. We expect that minimally broken invariance may explain cosmological observations. The purpose of including radiation is that it might cause a deviation in the Hubble constant problem.

 For radiation, the term $T^{\lambda}_{\lambda}=0$, and also $T_{\chi}=0$ since there is no direct coupling to the vector field with field $\chi$   \cite{Jain:2012gc}. The energy densities corresponding to matter and radiation still hold the same proportionality: $\rho_m \propto \chi^{-3}$ and $\rho_R \propto \chi^{-4}$ respectively, as in standard general relativity \cite{Jain:2012gc}.  Using Eqs. \ref{FEQ1} and \ref{FEQ2}, we obtain,
 \begin{equation}
  (\xi-2)\left(\frac{d \chi}{d \eta}\right)^2 - 2 \chi \frac{d^2 \chi}{d \eta^2} = \frac{\kappa}{2} \chi^4 \rho_{m} + \frac{2 \kappa}{3} \chi^4 \rho_{R}.
 \end{equation}
Here, we have used $T^{R}_{00}= \chi^4 \rho_R$. The equation of motion of $\chi$ gives,
\begin{equation}
 2 \xi \frac{d^2 \chi}{d \eta^2} = \kappa \chi^3 \rho_m + \kappa \Big[(\xi-2)\Lambda \chi^{\xi-3} + \Lambda_1 \Big]. \label{EQchi}
\end{equation}
Integrating Eq. (\ref{EQchi}), we obtain,
\begin{equation}
 \left(\frac{d \chi}{d\eta}\right)^2 = \frac{\kappa (\rho_{0m} + \Lambda_1) \chi}{\xi} + \frac{\kappa \Lambda}{\xi} \chi^{\xi -2} + C_0,
\end{equation}
where, $C_0$ is the integration constant. For a consistent solution, we get,
\begin{equation}
 \rho_{0m} = \frac{2 (\xi-3) \Lambda_1}{(6-\xi)},
\end{equation}
and 
\begin{equation}
  C_0 = \frac{2 \kappa \rho_{0R} }{3 (\xi-2)}.
 \end{equation}
Now, we consider $\chi= a(t)$, where $a(t)$ is the scale factor of the universe. Changing the conformal time ``$\eta$'' to the cosmological time ``$t$'' using $\chi d\eta = a(t) d\eta =dt$ leads to
\begin{equation}
 H(a)^2 = \frac{\kappa (\rho_{0m} + \Lambda_1)}{\xi} a^{-3} + \frac{\kappa \Lambda}{\xi} a^{\xi -6} +  \frac{2 \kappa \rho_{0R} }{3 (\xi-2)}a^{-4},
\end{equation}
and simplifying, we obtain,
\begin{eqnarray}
 H(a) &=& \frac{H_0}{\left(1+ (\rho_{0m} \xi)/(2 \Lambda (
 \xi-3)) + (2 \xi \rho_{0R})/(3 \Lambda (\xi-2))\right)^{1/2}} \nonumber \\  &&\times \  a^{\xi/2 -3} \Big[1+ \frac{\rho_{0m} \xi}{2 \Lambda (\xi-3)} a^{3-\xi} + \frac{2 \xi \rho_{0R}}{(3 \Lambda (\xi-2))} a^{2-\xi}\Big]^{1/2}, \label{Hub1}
\end{eqnarray}
Here, $\rho_{0m}$ represent the current energy density of ordinary matter and CDM together, $\rho_{0R}$ represent the current energy density of photons and neutrinos together, and $H_0$ is given by
\begin{eqnarray}
 H_0 = \sqrt{\frac{\kappa \Lambda}{\xi} \Big[1+ (\rho_{0m} \xi)/(2 \Lambda (
 \xi-3)) + (2 \xi \rho_{0R})/(3 \Lambda (\xi-2))\Big]}.
\end{eqnarray}
\section{Methodology and Data sets}
We will perform our analyses with the Hubble relation given in Eq. (\ref{Hub1}), which can be written in standard notation as  
\begin{eqnarray}
    H(z,\Omega_m,H_0,\xi) = H_0 (1+z)^{3-\xi/2} \sqrt{\left(\frac{3}{\xi-3}\Omega^0_m (1+z)^{\xi-3}+\frac{4}{\xi-2}\Omega^0_R (1+z)^{\xi-2}+\frac{6}{\xi}\Omega_\Lambda\right)},
    \label{Hz}
\end{eqnarray}
where, for each component, we define $\Omega_i^0=\rho_i^0/\rho_{cr}$ with $\rho_{cr}$ is critical energy density of the universe. For $\Lambda$CDM model we set $\xi=6$.
In the unimodular gravity model, we set ($H_0, \Omega_m, \xi$) as free parameters. We use the observational data sets including the Supernovae (SN), measurements of the BAO, and observational H(z) data obtained from the differential age method (DA) to constrain our parameters. The MCMC exploration of the model parameter space is carried out using the Python ensemble sampling toolkit \emph{emcee} \cite{emcee}.

The most recent data sample of Supernovae of type Ia, containing 1048 data points, was released as Pantheon dataset \cite{Scolnic:2017caz}, spanning the redshift range $0.01<z< 2.26$. These observations provide the apparent magnitude m(z) of the supernovae at peak brightness. The resulting apparent magnitude $m(z)$ is related to the luminosity distance $d_L(z)$ as,
\begin{equation}
    m_{th}(z,m_0,\Omega_m, \xi)= m_0(M,H_0) + 5 \log_{10} [D_L(z)].
\end{equation}
Here,
\begin{equation}
     D_L (z)= (1+z) \int_0^z \frac {H_0}{H(z',\Omega_i, \xi)}dz'
     \label{dl}
\end{equation}
is the Hubble free luminosity distance, defined as ($ D_L= H_0 d_L /c$) and ${m_0}$ is the marginalized nuisance parameter which depends on the absolute magnitude $M$ and the
present Hubble parameter $H_0$ as,
\begin{eqnarray}
  m_0(M,H_0) &=& M + 5 \log_{10}\left(\frac{c\; H_0^{-1}}{Mpc}\right) + 25 \nonumber \\
&=& M-5 \log_{10}H_0+52.38 .
\label{m0}   
\end{eqnarray}
The parameter $M$ represents the absolute magnitude of SN, which is assumed to be constant after including all corrections in $m(z)$ and $H_0$ is defined in the unit $\mbox{Km s}^{-1} \mbox{Mpc}^{-1}$. The Pantheon dataset gives the values of $m_{obs}$ in the redshift range $0.01<z< 2.26$. 
The parameters of the model are determined by minimizing the likelihood function $\mathcal L$ satisfying, $\ln \mathcal L = -\frac{1}{2}\chi^2_{SN}$, where
\begin{equation}
    \chi^2_{SN}=\sum_{i,j=1}^{1048} \Delta m(z_i,m_0,\Omega_m, \xi)^T C_{i,j}^{-1}\Delta m(z_j,m_0,\Omega_m, \xi).
\end{equation}
Here $C_{ij}$ is the full covariance matrix, and $\Delta m(z_i,m_0,\Omega_m, \xi) =  m_{obs}(z_i)-m_{th}(z_i,m_0,\Omega_m, \xi)$. The parameter $m_0$ is a nuisance parameter. After marginalizing $m_0$, the chi-square function is written as \cite{Nesseris:2005ur},
\begin{equation} {\chi}^2_{SN}(\Omega_m, \xi)=A(\Omega_m, \xi)-
\frac{B(\Omega_m, \xi)^2}{C} ,
\label{SN:chi2}
\end{equation}
where,
\begin{eqnarray}
A(\Omega_m, \xi)&=&\sum_{i,j=1}^{1048} \Delta m(z_i,m_0=0,\Omega_m, \xi)^T C_{i,j}^{-1}\Delta m(z_j,m_0=0,\Omega_m, \xi), \\
B(\Omega_m, \xi)&=&\sum_{i,j=1}^{1048} \Delta m(z_i,m_0,\Omega_m, \xi)^T C_{i,j}^{-1} , \\
C&=&\sum_{i,j=1}^{1048}C^{-1}_{ij}.
\end{eqnarray}

The SN data sets alone cannot constrain $H_0$, so we also consider the 31 observed Hubble data points from the differential age (DA) method \cite{Jimenez:2001gg} to constrain the value of $H_0$. The quantity measured in the differential age method is related to the Hubble parameter, 
\begin{equation}
    H(z)=-\frac{1}{1+z} \frac{dz}{dt}.
\end{equation}
This method can be used to find the Hubble constant $H_0$. Table \ref{tab:DA} shows the 31 points of $H(z)$ data given by differential age method \cite{Zhang:2020uan }.

\begin{table}[h!]
    \centering
     \caption{The 31 observational data points of $H(z)$ obtained from the differential age method.}
     \vspace{2em}
    \label{tab:DA}
    
    \begin{tabular}{|l c c c c c c c |}
        \hline
        \hline 
        $z$ & $H(z)$ & $\sigma_H$ & Ref. & $z$ & $H(z)$ & $\sigma_H$ & Ref.\\
        \hline
        0.09  & 69 &  12 & \cite{Jimenez:2003iv} & 0.3802 & 83 & 13.5 & \cite{Moresco:2016mzx} \\
        0.07 & 69.0 & 19.6 & \cite{Zhang:2012mp} & 0.4004 & 77 & 10.2 & \cite{Moresco:2016mzx}\\
        0.12 & 68.6 & 26.2 & \cite{Zhang:2012mp} & 0.4247 & 87.1 &  11.2 & \cite{Moresco:2016mzx}\\ 
        0.20 & 72.9 & 29.6 & \cite{Zhang:2012mp} & 0.4497 & 92.8 & 12.9 & \cite{Moresco:2016mzx}\\
        0.28 & 88.8 & 36.6 & \cite{Zhang:2012mp} &  0.4783 & 80.9 & 9 & \cite{Moresco:2016mzx}\\
        0.17 & 83 &  8  & \cite{Simon:2004tf} & 0.47 & 89 & 23 & \cite{Ratsimbazafy:2017vga} \\
        0.27 & 77 &  14 & \cite{Simon:2004tf}& 0.48 & 97 & 62 & \cite{Stern:2009ep} \\
        0.4  & 95  &  17 & \cite{Simon:2004tf}&  0.88 & 90 & 40 & \cite{Stern:2009ep}\\
        0.9  & 117 &  23 & \cite{Simon:2004tf}&  1.3  & 168 & 17 & \cite{Simon:2004tf}\\
        0.1791 & 75 & 4  & \cite{Moresco:2012jh}   &  1.43 & 177 & 18 & \cite{Simon:2004tf}\\
        0.1993 & 75 & 5  & \cite{Moresco:2012jh} &  1.53 & 140 & 14 & \cite{Simon:2004tf}\\
        0.3519 & 83 & 14 & \cite{Moresco:2012jh} &  1.75 & 202 & 40 & \cite{Simon:2004tf}\\
        0.5929 & 104 & 13 & \cite{Moresco:2012jh} & 1.037  & 154 & 20 & \cite{Moresco:2012jh}\\
        0.6797 & 92 & 8  & \cite{Moresco:2012jh} &  1.363  & 160 & 33.6  & \cite{Moresco:2015cya} \\ 
        0.7812 & 105 & 12 & \cite{Moresco:2012jh} & 1.965  & 186.5 &  50.4  & \cite{Moresco:2015cya}\\

        0.8754 & 125 & 17 & \cite{Moresco:2012jh} & & & &\\     
        
        \hline
        \end{tabular}

\end{table}
We perform the joint analysis of the SN and DA data sets by minimizing the chi-square function, defined as
\begin{equation}
    \chi^2 = \chi^2_{SN} +\chi^2_{DA},
\end{equation}
where $\chi^2_{SN}$ is defined in Eq. (\ref{SN:chi2}) and $\chi^2_{DA}$ is given by,
\begin{equation}
 \chi^2_{DA} =\sum \frac{(H(z)-H(z,\Omega_m,H_0,\xi))^2}{\sigma_H^2} .  
\end{equation}

We further estimate the model parameters using the BAO data set. We use the observational data sets of the BAO measurements including Galaxy BAO and $L{y\alpha}$ BAO (eBOSS). BAO studies along the line of sight measure the combination $H(z)r_d$, whereas investigations of the BAO feature in the transverse direction offer a value of $D_M(z)/r_d$ and $D_V(z)/r_d$, where $r_d \equiv r_s(z_d)$  is the comoving size of the sound horizon at the baryon drag epoch $(z_d)$ \cite{Eisenstein:1997ik}. Table (\ref{tab:BAO}) contains a list of the data sets we used.
\begin{table}[h]
\centering
\caption{BAO data measurements we used in our analysis.
$D_M$, $D_V$, and $r_d$ are in units of Mpc, while $H(z)$ is in units of km s$^{-1}$ Mpc$^{-1}$.}
\vspace{1em}
\label{tab:BAO}
\begin{tabular}{cccccc}
  \hline
  \hline
    Survey & z$_{ef}$ & Measurement & Observation & error & Reference \\
   \hline
6dFGS      &     0.106   &    $D_V/r_d$     &     	2.9762     &   	0.1329  &  \cite{Lemos:2018}\\
SDSS MGS     &   0.15    &    $D_V/r_d$    	&	4.4657    &    	0.1681   &	 \cite{Ross:2014}\\
BOSS DR12   &   	0.38    &   $D_M \times r_{d, fid}/r_d$ 	&	1518      &  	20  &	\cite{BOSS:2016}\\
BOSS DR12   &   	0.38    &   $H \times r_d/r_{d,fid}$    &	81.5	&	1.7  	&	\cite{BOSS:2016}\\
BOSS DR12   &    0.51     &   $D_M \times r_{d,fid}/r_d $ 	& 1977      &      24  &	 \cite{BOSS:2016}\\
BOSS DR12     &  0.51      &  $ H \times r_d/r_{d,fid} $    &	90.5	&	1.7    &   	\cite{BOSS:2016}\\ 
BOSS DR12    &   0.61      &  $ D_M \times r_{d,fid}/r_d$ &	2283       & 	28      &   \cite{BOSS:2016}\\
BOSS DR12     &  0.61      &  $ H \times r_d/r_{d,fid} $  & 	97.3	&	1.8   	&	\cite{BOSS:2016}\\
BOSS DR14     &  0.72    &   $ D_V/r_d  $  	&	16.08472   &     0.41278 &   \cite{Bautista:2017}\\
eBOSS QSO	& 0.978      & $ D_A \times r_{d,fid}/r_d$ &	1586.18		& 284.93	&	 \cite{Zhao:2018}\\
eBOSS QSO	& 0.978     & $ H \times r_d/r_{d,fid} $	 &113.72	&	14.63 	&	 \cite{Zhao:2018}\\
eBOSS QSO  &	1.23     & $  D_A \times r_{d,fid}/r_d $	&1769.08	&	159.67 	&	 \cite{Zhao:2018}\\
eBOSS QSO &	1.23      & $ H \times r_d/r_{d,fid}  $	 & 131.44	&	12.42 	&	 \cite{Zhao:2018}\\
eBOSS QSO & 	1.526  &   $  D_A \times r_{d,fid}/r_d 	$ &1768.77	&	96.59   & 	 \cite{Zhao:2018}\\
eBOSS QSO &	1.526      & $ H \times r_d/r_{d,fid} $  &	148.11 &	12.75 	&	 \cite{Zhao:2018}\\
eBOSS QSO  & 	1.944   &  $  D_A \times r_{d,fid}/r_d $ &	1807.98		&146.46 &	 \cite{Zhao:2018}\\
eBOSS QSO &	1.944     & $ H \times r_d/r_{d,fid} $  &	172.63		&14.79 	&	 \cite{Zhao:2018}\\
eBOSS Ly  &	2.34   &   $  D_M/r_d $ &	37.41	&	1.86 	&	 \cite{Blomqvist:2019}\\
eBOSS Ly &	2.34       & $ _H/r_d	$ &	8.86	&	0.29 		& \cite{Blomqvist:2019}\\
eBoss QSOxLy &  2.35   &   $  D_M/r_d $ &  		36.3   &   	1.8 &		 \cite{Blomqvist:2019}\\
eBoss QSOxLy  & 2.35    &   $ D_H/r_d  $ & 		9.20  &    	0.36 	&	\cite{Blomqvist:2019}\\
eBoss combined & 2.34     &  $ D_M/r_d  $  &		37    &  	1.3 	&	 \cite{Blomqvist:2019}\\
eBoss combined  & 2.34     & $  D_H/r_d  $  &		9.00  &    	0.22 	&	\cite{Blomqvist:2019}\\
  \hline
\end{tabular}
\end{table}

The comoving angular diameter distance $D_M(z)$ and volume averaged scale $D_V(z)$ are related to $H(z)$ as
\begin{eqnarray}
    D_M(z)&=& c \int_0^{z'} \frac{dz}{H(z')},\\
    D_V(z)&=& \left(z D_H(z) D_M^2(z)\right)^{1/3},\\
    D_A(z)& =& \frac{D_M}{(1+z)}.
\end{eqnarray}

The comoving sound horizon is given by
\begin{equation}
    r_d=\int_{z_d}^\infty \frac{c_s(z)}{H(z)} dz ,
    \label{rd}
\end{equation}
where $c_s(z)= c[3+\frac{9}{4} \rho_b(z)/\rho_\gamma(z)]^{-1/2} $ is the speed of sound in the baryon-photon fluid, with $\rho_b(z)$ and $\rho_\gamma(z)$ being the baryon and photon densities respectively, and $z_d$ being the redshift at the drag epoch. The reference point used to calibrate the BAO observations is the sound horizon $r_d$, also known as the standard ruler of BAO observations. From BAO data, we can only constrain the combination of $H_0$ and sound horizon $r_d$. In order to approximate the $H_0$ from BAO data, we use the analytic approximation of $z_d$ from Ref \cite{Wayne-Hu} and take $\Omega_b h^2$ from Planck 2018 results ($\Omega_b h^2 =0.0224$) \cite{Planck:2018vyg}.

 We also combine the BAO data with the DA and SN data sets and perform the joint analysis to explore the parameter space. We minimize the $\chi^2$ function, defined as
\begin{equation}
    \chi^2 =\chi^2_{SN} +\chi^2_{DA}+\chi^2_{BAO}.
\end{equation}

\vspace{2.0cm}
\section{Results and Discussion}
\begin{figure}[ht!]
\centering
   \includegraphics[width=8cm]{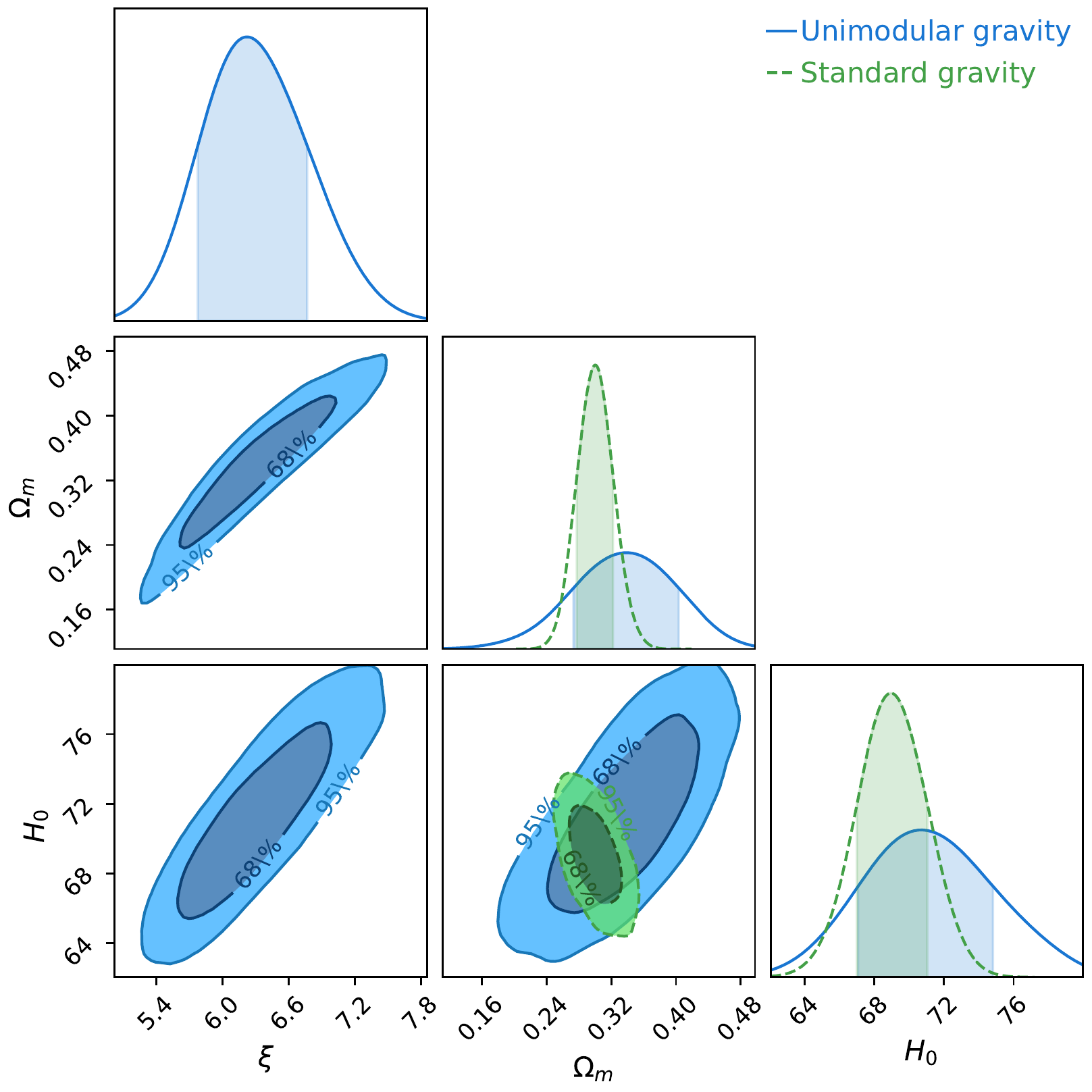}
   \caption{2D contour plot and  1D marginalized posterior distributions of model parameters $\xi$, $\Omega_m$, and $H_0$ in unimodular gravity model and standard ($\Lambda$CDM) model using the SN+DA data sets.}
   \label{SN_DA_full}
\end{figure}

In unimodular gravity model, we set the uniform prior on all three parameters ($\Omega_m, H_0,\xi$) in the range $\Omega_m \in [0,0.5], \, H_0 \in [60,80]$, and $\xi \in [5,8]$ and perform the MCMC analysis of joint SN+DA dataset. The mean value of the parameters we obtained are $\Omega_m=0.338 \pm 0.065$, $\xi=6.23 \pm 0.50$, and $H_0 = 70.7^{4.1}_{-3.6} \,\mbox{Km s}^{-1} \mbox{Mpc}^{-1}$. For the standard ($\Lambda$CDM) model ($\xi=6$), the mean value of the parameters obtained by joint analysis are $H_0 = 69.0 \pm 2.1 $ Km s$^{-1}$ Mpc$^{-1}$ and $\Omega_m =0.30 \pm 0.02$. In Fig. \ref{SN_DA_full} we show the results for unimodular gravity and standard gravity model using the SN+DA data sets.  \\

We note that while unimodular gravity theory ($\xi \neq 6$) can fit the low redshift SN+DA data sets, the constraint on $\Omega_m$ and $H_0$ are not as tight as in the case of standard gravity $\Lambda$CDM model.

We next assume the unimodular gravity with ($\xi=6.23$) as the underlying theory of gravity, set the uniform prior on $\Omega_m$ and $H_0$ and perform the MCMC analysis using BAO data set. We obtained the mean value of the parameters as, $\Omega_m=0.295 \pm 0.015$, and $H_0 = 69.24 \pm 0.90 \,\mbox{Km s}^{-1} \mbox{Mpc}^{-1}$. If we put these values in Eq. (\ref{rd}), we find that the sound horizon $r_d$ in unimodular gravity ($\xi=6.23$) is $152.5$ Mpc. The mean values of the parameters we found for the standard ($\Lambda$CDM) model ($\xi=6$) are, $\Omega_m =0.288 \pm 0.018$, $H_0 = 67.7 \pm 1.1 $ Km s$^{-1}$ Mpc$^{-1}$, and in this case $r_d=150.2$ Mpc.

\begin{figure}[ht!]
\centering
   \includegraphics[width=8cm]{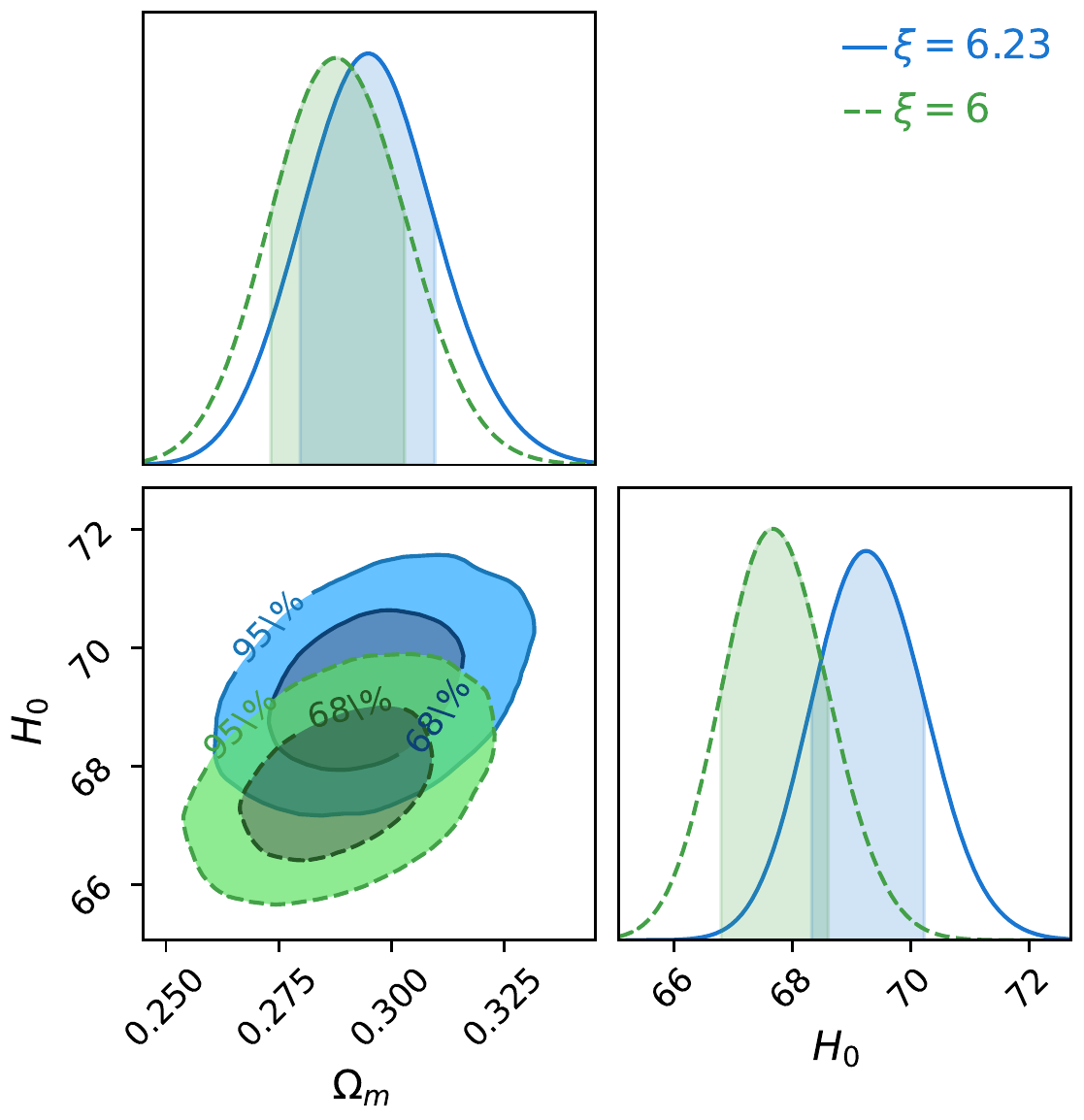}
   \caption{2D contour plot and 1D posterior distributions of model parameters $\Omega_m$ and $H_0$ in unimodular gravity model ($\xi=6.23$) and standard ($\Lambda$CDM) model using the BAO data sets.}
   \label{BAO:lcdm-UG}
\end{figure}

We note that while the mean value of $\Omega_m$ in unimodular gravity with BAO data set is consistent with the standard gravity model, the mean value of $H_0$ is having more than $1\sigma$ difference with the standard $\Lambda$CDM model. Also, the value of $r_d$ differs in both models. In Fig. \ref{BAO:lcdm-UG}, we show the comparison of the unimodular gravity model with the standard gravity model using BAO data sets.

\begin{figure}[h!]
\centering
   \includegraphics[width=8cm]{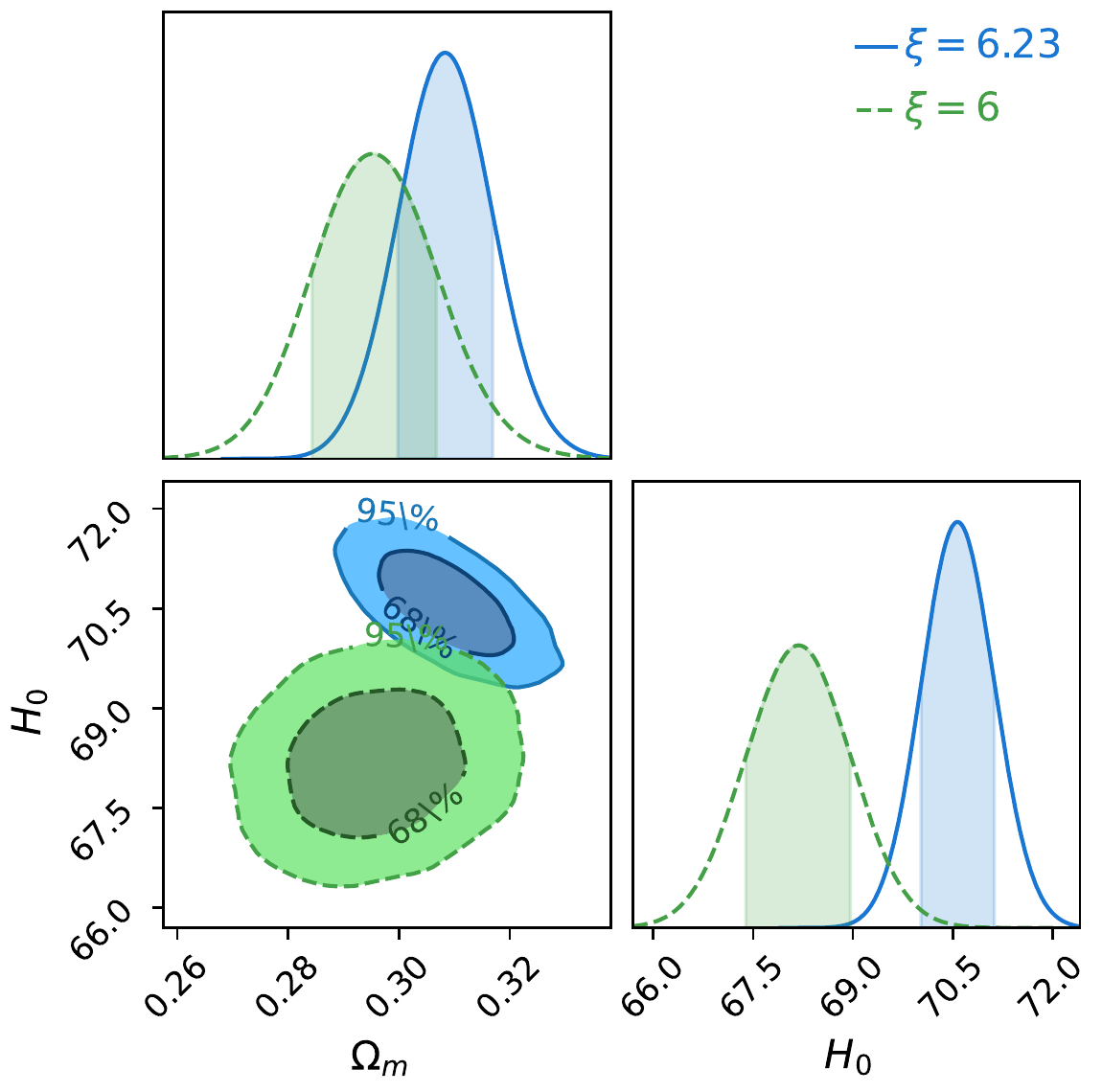}
   \caption{2D contour plot and 1D posterior distributions of model parameters $\Omega_m$ and $H_0$ in unimodular gravity model ($\xi=6.23$) and standard ($\Lambda$CDM) model using the BAO+SN+DA data sets.}
   \label{full_posterior}
\end{figure}

Combining BAO data with the SN+DA data sets we get $\Omega_m=0.308 \pm 0.008$ and $H_0 =70.57 \pm 0.56 \,\mbox{Km s}^{-1} \mbox{Mpc}^{-1}$ as best fit parameters for unimodular gravity model and  $\Omega_m=0.295 \pm 0.011$ and $H_0 =68.17 \pm 0.80 \, \mbox{Km s}^{-1} \mbox{Mpc}^{-1}$ for standard ($\Lambda$CDM) model. We find that the value of $\Omega_m$ in the unimodular gravity model ($\xi=6.23$) using the SN+DA+BAO data set is consistent with the Planck 2018 results. We also find the value of sound horizon in unimodular gravity $ r_d=149.4$ Mpc, whereas from Planck 2018 results $r_d\sim 147$ Mpc. The results of both models are shown in Fig. \ref{full_posterior}. In Table \ref{tab:result}, we listed the best fit value of parameters inferred from BAO, DA, and SN data sets in the standard ($\Lambda$CDM) model and unimodular gravity model.

\begin{table}[!ht]
\centering
\caption{The results in standard ($\Lambda$CDM) model and unimodular gravity model using BAO, DA, and SN data sets.}
\label{tab:result}
\vspace{1em}
\begin{tabular}{cccc}
  \hline
  \hline
   Data sets & Parameter  & Standard gravity & Unimodular gravity \\
  \hline
   & $H_0$  & $69.0 \pm 2.1$  & $70.7 \pm 4.1$ \\
  SN+DA &  $\xi$ & $6 $ & $6.23 \pm 0.5$   \\
 & $\Omega_m$ & $0.3 \pm 0.02$ & $0.338 \pm 0.065$  \\
 \hline
 & $H_0$  & $67.7 \pm 1.1$  & $69.24 \pm 0.9 $ \\
 BAO &  $\xi$ & $6 $ & $6.23 \pm 0.5$   \\
 & $\Omega_m$ & $0.288 \pm 0.018$ & $0.295 \pm 0.015$  \\
 & $r_d$ & $150.2$ & $152.5$  \\
  \hline
   & $H_0$  & $ 68.17 \pm 0.8$  & $70.57 \pm 0.56 $ \\
 BAO+SN+DA &  $\xi$ & $6 $ & $6.23 \pm 0.5$   \\
 & $\Omega_m$ & $0.295 \pm 0.011$ & $0.308 \pm 0.008$  \\
 & $r_d$ & $148.8$ & $149.4$  \\
 \hline
\end{tabular}
\end{table}

\begin{figure}[ht]
     \begin{subfigure}[b]{0.45\textwidth}
         \centering
         \includegraphics[width=7cm]{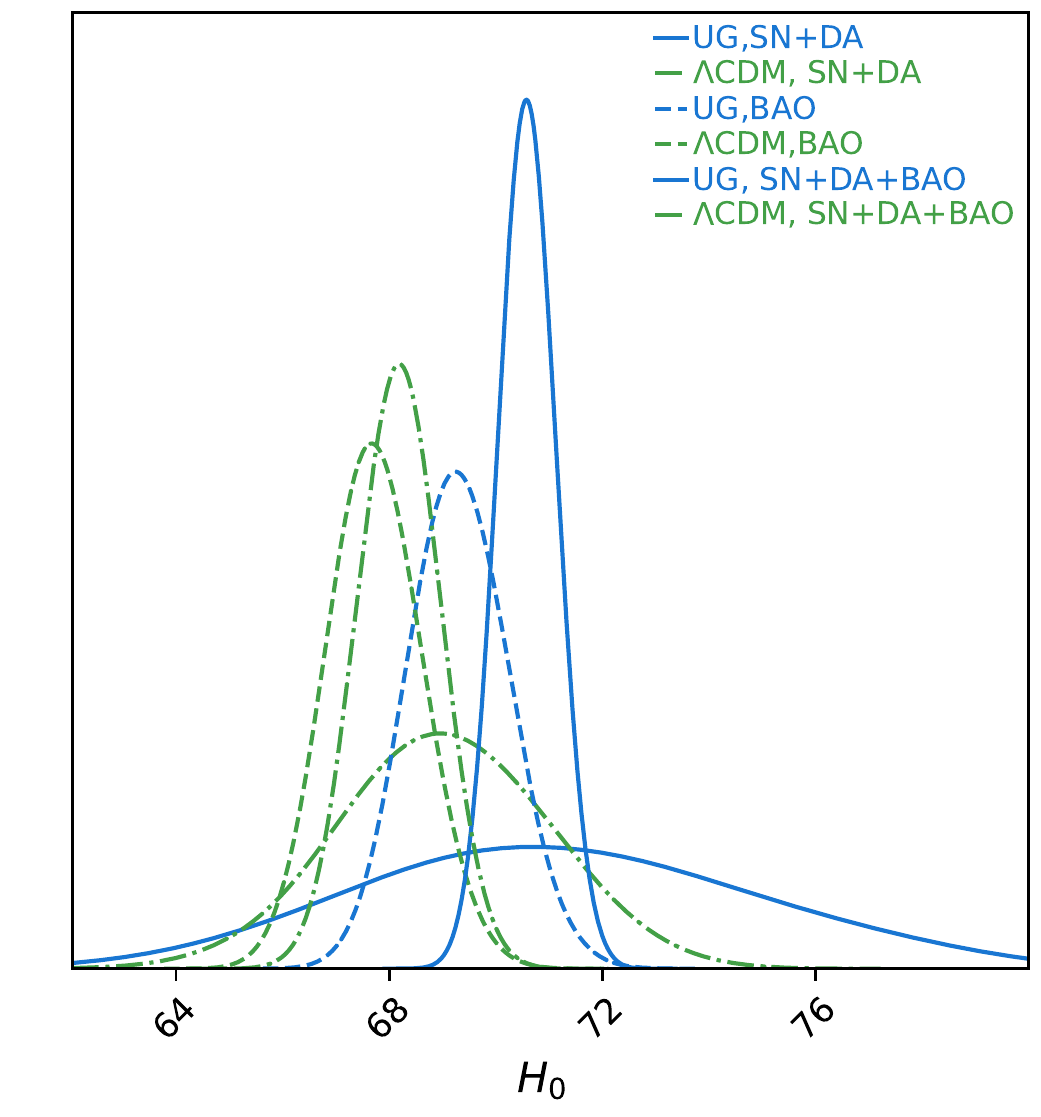}
     \end{subfigure}
     \begin{subfigure}[b]{0.45\textwidth}
         \centering
         \includegraphics[width=8.5cm]{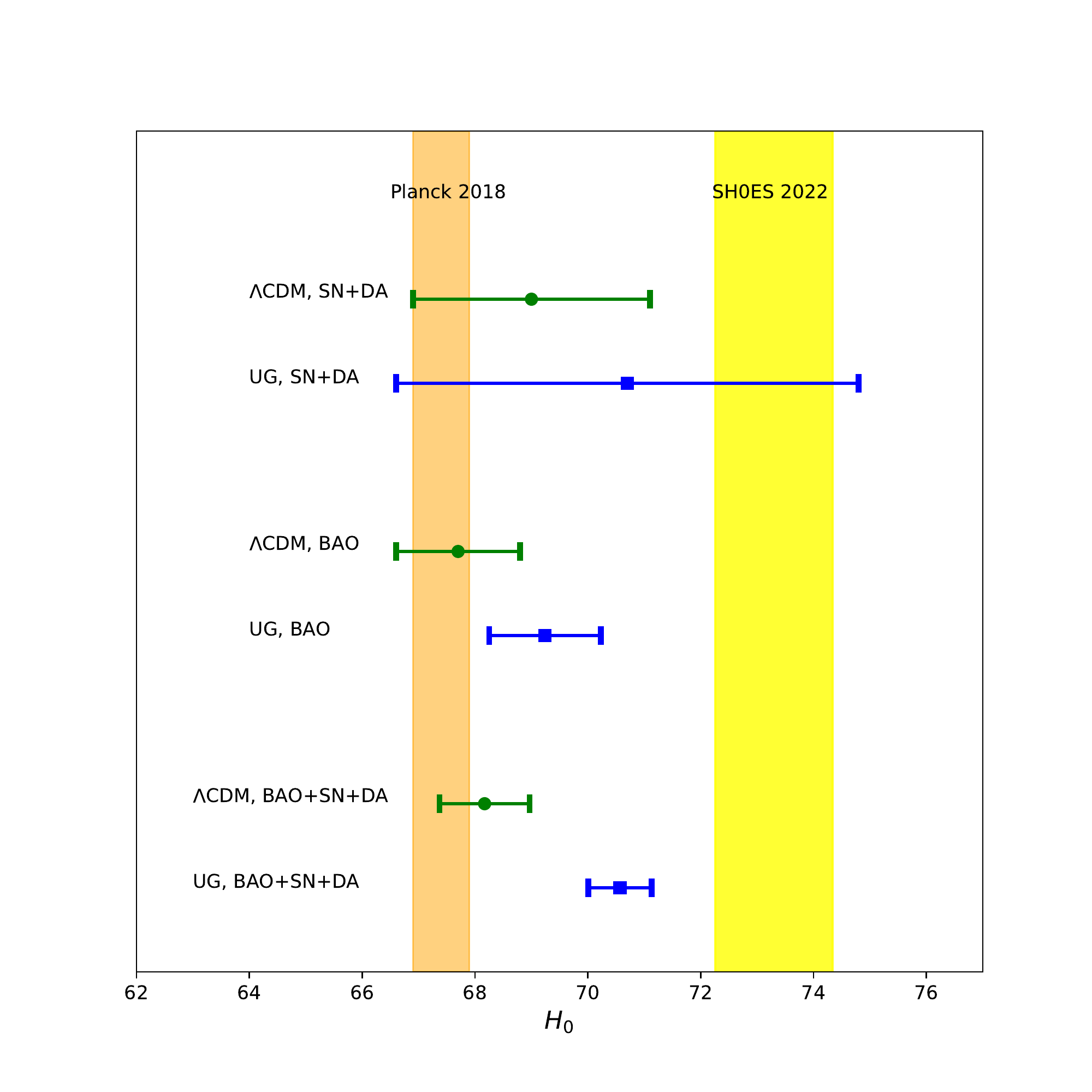}
     \end{subfigure}
      \caption{1D posterion distribution and corresponding median along with 1$\sigma$ error band plot of $H_0$ for UG and standard ($\Lambda$CDM) model in comparison with Planck 2018 and SH0ES 2022 results.}
      \label{H0_comparison}
\end{figure}

In order to study the impact of unimodular gravity on Hubble tension problem, in Fig. \ref{H0_comparison}, we plotted the 1D posterior distribution of $H_0$ and the corresponding median along with the 1$\sigma$ error band plot for unimodular gravity and standard ($\Lambda$CDM) model. Here we also compare our results with the Planck 2018 and SH0ES 2022 estimates of $H_0$.

From these results, one can note that the value of $H_0$ depends on $\xi$. Here we have estimated the $\xi=6.23$ from SN+DA data sets and use this value consistently in other data analyses. The mean value of $H_0$ obtained from SN+DA data almost remains unchanged after including the BAO data, but the constraint becomes more stringent. Furthermore, the SH0ES 2018 result differ by $\sim 2.5\sigma$ from our results.

\section{Conclusion} In this paper, we have discussed the cosmological implication of unimodular theory of gravity, which is an alternative approach to the general theory of relativity. The decomposition of full metric in a scalar field and unimodular metric leads to decomposition in affine connection, Ricci tensor, and Ricci scalar. General coordinate invariance is broken by introducing a parameter $\xi$. We considered an unimodular gravity model  with broken general coordinate invariance and focused on estimating the cosmological parameters of the current universe in this model.
We found that the UG model can fit the latest supernovae data, combined with observed H(z) data from the differential age  method, with $\xi \approx 6.23$, which is very close to the general relativity ($\xi=6$). We used the value of $\xi=6.23$ consistently and further estimated the cosmological parameters using the BAO data set and joint BAO+SN+DA data sets. We also estimated the cosmological parameters in the standard $\Lambda$CDM model for comparison. We found that in the UG model using SN+DA data sets, Hubble constant $H_0$ is $ 70.7 \pm 4.1 \ \mbox{Km} \ \mbox{Sec}^{-1} \ \mbox{Mpc}^{-1}$, and using BAO constraint, $H_0$  is $69.24 \pm 0.90 \ \mbox{Km} \ \mbox{Sec}^{-1}  \mbox{Mpc}^{-1}$. We observe that in unimodular gravity, using SN+DA and BAO data sets, the mean value of $H_0$ shifted to the higher value, reducing tension with the SH0ES 2022 result. However, from SN+DA in unimodular gravity, we obtain a larger dispersion in $H_0$ than that of standard gravity. We have finally combined all the data (BAO+SN+DA) and estimated the Hubble constant as $70.57 \pm 0.56 \ \mbox{Km} \ \mbox{Sec}^{-1} \ \mbox{Mpc}^{-1}$ and $68.17 \pm 0.80  \ \mbox{Km} \ \mbox{Sec}^{-1} \ \mbox{Mpc}^{-1}$ in unimodular gravity and standard gravity, respectively. A comparison with the model used in Ref. \cite{Jain:2012gc} is necessary. In that model, only generalised cosmological constant or generalised non-relativistic matter is considered, whereas in our model, cold dark matter, generalised cosmological constant, and radiation are all taken into account. Therefore, our model provides a more satisfactory fit to the data. It could be a reliable model that consists of cold dark matter and the cosmological constant similar to the $\Lambda$CDM of standard gravity.
 
\section*{Acknowledgements}
We thank Prof. Pankaj Jain for valuable suggestions and comments on the draft.

\bibliographystyle{JHEP}
 \bibliography{Unimodular_ref}

\end{document}